\documentclass[english,aps,reprint,twocolumn,superscriptaddress]{revtex4-1}
\usepackage[T1]{fontenc}
\usepackage[latin9]{inputenc}
\usepackage{amstext}
\usepackage{amssymb}
\usepackage{esint}
\usepackage{graphicx}
\usepackage{color}
\usepackage{babel}
\usepackage{epstopdf}
\begin{document}

\title{Multiphoton resonance band and Bloch-Siegert shift in a bichromatically driven qubit}

\author{Yiying Yan}\email{yiyingyan@zust.edu.cn}
\affiliation{Department of Physics, School of Science, Zhejiang University of Science and Technology, Hangzhou 310023, China}
\author{Zhiguo L\"{u}}\email{zglv@sjtu.edu.cn}
\affiliation{Key Laboratory of Artificial Structures and Quantum Control
(Ministry of Education), School of Physics and Astronomy,
Shanghai Jiao Tong University, Shanghai 200240, China}
\author{Lipeng Chen}
\affiliation{Zhejiang Laboratory, Hangzhou 311100, China}
\author{Hang Zheng}
\affiliation{Key Laboratory of Artificial Structures and Quantum Control
(Ministry of Education), School of Physics and Astronomy,
Shanghai Jiao Tong University, Shanghai 200240, China}
\date{\today}

\begin{abstract}
We study the resonance and dynamics of a qubit exposed to a strong aperiodic bichromatic field by using a periodic counter-rotating hybridized rotating wave (CHRW) Hamiltonian, which is derived from the original Hamiltonian with the unitary transformations under a reasonable approximation and enables the application of the Floquet theory. It is found that the consistency between the CHRW results and numerically exact generalized-Floquet-theory (GFT) results in the valid regime of the former while the widely used rotating-wave approximation (RWA) breaks down. We illustrate that the resonance exhibits band structure and the Bloch-Siegert shifts induced by the counter-rotating couplings of the bichromatic field become notable at the multiphoton resonance band. In addition, the CHRW method is found to have a great advantage of efficiency over the GFT approach particularly in the low beat-frequency case where the latter converges very slowly. The present CHRW method provides a highly efficient way to calculate the resonance frequency incorporating the Bloch-Siegert shift and provides insights into the effects of the counter-rotating couplings of the bichromatic field in the strong-driving regimes.
\end{abstract}

\maketitle

\section{Introduction}
Quantum systems driven by external fields have attracted great attention in physics and have been studied extensively in theory and experiments~\cite{Bloch_1940,Shirley_1965,Grifoni_1998,Saiko_2018,Bonifacio_2020,Ying_2020,M_rzac_2021,Macovei_2022,Ying_2021,Downing_2022}.
In recent years, this topic has been renewed in the context of artificial atoms such as semiconductor quantum dot~\cite{Wismer_2016,Koski_2018}, superconducting circuit~\cite{Tuorila_2010,Yoshihara_2014,Deng_2015,Pietik_inen_2017,Buks_2020,Magazz__2021,Chen_2021}, nitrogen-vacancy center~\cite{Shu_2018,Wang_2021}, etc. Owing to the controllability of these artificial atoms, it has been realized experimentally that the strong or ultrastrong interaction between the quantum system and external field,  where the driving strength becomes comparable with or even exceeds the transition frequency of the quantum system~\cite{Tuorila_2010,Yoshihara_2014}. The strong driving results in the invalidity of the widely used rotating-wave approximation (RWA) and interesting effects of counter-rotating couplings.
As is well-known, in the Rabi model which describes a two-level system (qubit) driven by a monochromatic field, the counter-rotating coupling is found to cause multiphoton quantum resonances~\cite{Shirley_1965,Aravind_1984} and the Bloch-Siegert shift which describes that the resonance frequency varies with the driving strength~\cite{Bloch_1940,Shirley_1965,Stenholm_1972,Hannaford_1973,Cohen_Tannoudji_1973,L__2012}.

The coherent interaction between a two-level system and a bichromatic field is an ever-green problem and has attracted considerable attentions~\cite{Guccione_Gush_1974,Ho_1984,Ruyten_1989,Ruyten_1992,Agarwal_1991,Ficek_1996,Pan_2017,Gustin_2021}, ranging from dynamics to fluorescence. There are rich multiphoton resonant processes such as that the two-level system absorbs $n+1$ photons from one component of the bichromatic field and emits $n$ photons of the other component with $n$ being an positive integer, which have been revealed by a number of theoretical studies based on the Green's function~\cite{Guccione_Gush_1974}, generalized Floquet theory (GFT)~\cite{Ho_1984}, and RWA~\cite{Ruyten_1989,Ruyten_1992}. If the RWA is used, it is possible to reduce the aperiodic Hamiltonian to a periodic one in a proper rotating frame and thus the Floquet theory can be applied. In references~\cite{Ruyten_1989,Ruyten_1992}, the resonance shifts have been illustrated with the combination of the Floquet theory and continued fraction in the absence of the counter-rotating coupling. The GFT allows a treatment beyond the RWA and converts the time-dependent problem into an eigenvalue problem of an infinite-size generalized Floquet matrix. However, it is difficult to be diagonalized exactly by analytical methods and one usually carries out perturbation calculation and numerical calculation~\cite{Chu_2004}. In reference~\cite{Ho_1984}, the perturbative analytical calculation has shown that there are Bloch-Siegert type resonance shifts when the counter-rotating couplings are taken into account. So far, most studies focus on the cases in which the beat frequency of bichromatic field is a significant fraction of the driving frequency. Few efforts have been devoted to the case in which the beat frequency is a vanishingly small fraction of the driving frequency. In the latter case, the convergence of the GFT approach may be problematic.

In this work, we postulate a periodic effective Hamiltonian to study the resonance incorporating the Bloch-Siegert shift and dynamics of the bichromatically driven qubit. The effective Hamiltonian is derived with the unitary transformations under a reasonable approximation which is valid in a strong-driving regime of interest and is referred to as the counter-rotating hybridized rotating wave (CHRW) Hamiltonian. Taking the advantage of the periodicity of the CHRW Hamiltonian, we are able to calculate the time-averaged and transient transition probabilities, and resonance frequencies by using the Floquet theory.  When the beat frequency of the bichromatic field is comparable with its frequencies, we compare the results calculated from the CHRW Hamiltonian with the RWA results and the numerically exact GFT results.
It turns out that in the valid regime the CHRW method is highly efficient and satisfactorily accurate in comparison with the GFT method. A comparison between the CHRW and RWA results reveals the Bloch-Siegert shifts induced by the counter-rotating couplings. We show that the Bloch-Siegert shifts become notable at the multiphoton resonance bands and are found to play an important role in the multiphoton dynamical processes under certain conditions. When the beat frequency of the bichromatic field is far smaller than its frequencies, we find that the GFT approach converges very slowly and thus we compare the transient transition probabilities of the CHRW and the Runge-Kutta (RK) methods to validate the former. It is found that the CHRW method is capable of efficiently providing accurate results and has a great advantage over the GFT. The Bloch-Siegert shift is also illustrated with the CHRW method in the low beat-frequency regimes. The present method provides a highly efficient way to calculate the resonance frequencies and insights into the effects of the counter-rotating couplings.

\section{Model and methodology}\label{sec:method}
\subsection{Unitary transformation}
The Hamiltonian describing a bichromatically driven qubit reads $(\hbar=1)$
\begin{equation}\label{eq:Ham}
H(t)=\frac{1}{2}\omega_{0}\sigma_{z}+\sum_{j=1}^{2}\frac{A_{j}}{2}\cos(\omega_{j}t+\phi_{j})\sigma_{x},
\end{equation}
where $\omega_{0}$ is the transition frequency of the qubit,
$\sigma_{\mu}$ $(\mu=x,y,z)$ is the Pauli matrix, $A_{j}$,
$\omega_{j}$, and $\phi_{j}$ ($j=1,2$) are the amplitude, frequency,
and phase of the $j$th component of the bichromatic field, respectively. We consider that $\omega_1$ and $\omega_2$ are incommensurate and thus $H(t)$ is aperiodic.
In the following, we derive a periodic effective Hamiltonian for the present aperiodic Hamiltonian.

We transform the Hamiltonian with the unitary transformation~\cite{L__2012}
\begin{eqnarray}
H^{\prime}(t) & = & {\rm e}^{S(t)}H{\rm e}^{-S(t)}-{\rm i}{\rm e}^{S(t)}\frac{d}{dt}{\rm e}^{-S(t)}\nonumber \\
 & = & \frac{\omega_{0}}{2}\left\{ \sigma_{z}\cos\left[\sum_{j=1}^{2}\frac{A_{j}}{\omega_{j}}\xi_{j}\sin(\omega_{j}t+\phi_{j})\right]\right.\nonumber\\
 &  &\left.+\sigma_{y}\sin\left[\sum_{j=1}^{2}\frac{A_{j}}{\omega_{j}}\xi_{j}\sin(\omega_{j}t+\phi_{j})\right]\right\} \nonumber \\
 &  & +\sum_{j=1}^{2}\frac{A_{j}}{2}(1-\xi_{j})\cos(\omega_{j}t+\phi_{j})\sigma_{x},
\end{eqnarray}
with the generator
\begin{equation}
S(t)={\rm i}\sum_{j=1}^{2}\frac{A_{j}}{2\omega_{j}}\xi_{j}\sin(\omega_{j}t+\phi_{j})\sigma_{x},
\end{equation}
where $\xi_{j}\in(0,1)$ are the parameters to be determined later.

Using the following expansions derived from the generating function of the Bessel functions of the first kind:
\begin{widetext}
\begin{eqnarray}
\cos\left(\sum_{j=1}^{2}z_{j}\sin\theta_{j}\right) & = & \sum_{k=1}^{\infty}\sum_{n=1}^{2k-1}2J_{n}(z_{1})J_{2k-n}(z_{2})\left\{ \cos[n\theta_{1}+(2k-n)\theta_{2}]\right.\nonumber\\
 &  &\left.+(-1)^{n}\cos[n\theta_{1}-(2k-n)\theta_{2}]\right\} \nonumber \\
 &  & +J_{0}(z_{1})J_{0}(z_{2})+2\sum_{n=1}^{\infty}[J_{2n}(z_{1})J_{0}(z_{2})\cos(2n\theta_{1})\nonumber\\
 &  &+J_{0}(z_{1})J_{2n}(z_{2})\cos(2n\theta_{2})],
\end{eqnarray}
\begin{eqnarray}
\sin\left(\sum_{j=1}^{2}z_{j}\sin\theta_{j}\right) & = & 2\sum_{k=1}^{\infty}\sum_{n=1}^{2k}J_{n}(z_{1})J_{2k+1-n}(z_{2})\left\{ \sin[n\theta_{1}+(2k+1-n)\theta_{2}]\right.\nonumber\\
 &  &\left.-(-1)^{n}\sin[n\theta_{1}-(2k+1-n)\theta_{2}]\right\} \nonumber \\
 &  & +2\sum_{n=1}^{\infty}\left\{ J_{2n-1}(z_{1})J_{0}(z_{2})\sin[(2n-1)\theta_{1}]\right.\nonumber\\
 &  &\left.+J_{0}(z_{1})J_{2n-1}(z_{2})\sin[(2n-1)\theta_{2}]\right\} ,
\end{eqnarray}
\end{widetext}
where $\theta_j=\omega_{j}t+\phi_j$, $z_{j}=\frac{A_{j}}{\omega_{j}}\xi_{j}$, and $J_{n}(z_{j})$
are the Bessel functions of the first kind,
we divide the transformed Hamiltonian into two parts according to
oscillating behaviors. The first part consists of the relatively
slow-varying terms: the time-independent term, the lowest beat-frequency term, and the single harmonic
terms, which reads
\begin{eqnarray}
H_{{\rm CHRW}}^{\prime}(t) & = & \frac{1}{2}\omega_{0}\{J_{0}(z_{1})J_{0}(z_{2})-2J_{1}(z_{1})J_{1}(z_{2})\nonumber\\
 &  &\times\cos[(\omega_{1}-\omega_{2})t+\phi_{1}-\phi_{2}]\}\sigma_{z}\nonumber \\
 &  & +\omega_{0}[J_{1}(z_{1})J_{0}(z_{2})\sin(\omega_{1}t+\phi_{1})\nonumber\\
 &  &+J_{0}(z_{1})J_{1}(z_{2})\sin(\omega_{2}t+\phi_{2})]\sigma_{y}\nonumber \\
 &  & +\sum_{j=1}^{2}\frac{A_{j}}{2}(1-\xi_{j})\cos(\omega_{j}t+\phi_{j})\sigma_{x}.
\end{eqnarray}
This part can be simplified via setting
\begin{equation}
\omega_{0}J_{1}(z_{1})J_{0}(z_{2})=\frac{A_{1}}{2}(1-\xi_{1}),
\end{equation}
\begin{equation}
\omega_{0}J_{0}(z_{1})J_{1}(z_{2})=\frac{A_{2}}{2}(1-\xi_{2}),
\end{equation}
which determine the values of $\xi_{j}$. By Taylor expansion and supposing that $A_1$ and $A_2$ have the similar order, one
readily finds that
\begin{eqnarray}
\xi_{1}&=&\frac{\omega_{1}}{\omega_{0}+\omega_{1}}\left[1+\frac{\omega_{0}A_{1}^{2}}{8(\omega_{0}+\omega_{1})^{3}}\left(1+2r^{2}\frac{(\omega_{0}+\omega_{1})^{2}}{(\omega_{0}+\omega_{2})^{2}}\right)\right]\nonumber\\
& &+O(A_{1}^{4}),
\end{eqnarray}
\begin{eqnarray}
\xi_{2}&=&\frac{\omega_{2}}{\omega_{0}+\omega_{2}}\left[1+\frac{\omega_{0}A_{1}^{2}}{8(\omega_{0}+\omega_{2})^{3}}\left(r^{2}+2\frac{(\omega_{0}+\omega_{2})^{2}}{(\omega_{0}+\omega_{1})^{2}}\right)\right]\nonumber\\
& &+O(A_{1}^{4}),
\end{eqnarray}
where we have defined the ratio
\begin{equation}
  r=A_{2}/A_{1}.
\end{equation}
Recall that $\sin(\omega t)\sigma_{y}+\cos(\omega t)\sigma_{x}={\rm e}^{{\rm i}\omega t}\sigma_{-}+{\rm e}^{-{\rm i}\omega t}\sigma_{+}$,
where $\sigma_{\pm}=(\sigma_{x}\pm i\sigma_{y})/2$. We can rewrite
$H_{{\rm CHRW}}^{\prime}(t)$ as
\begin{eqnarray}
H_{{\rm CHRW}}^{\prime}(t) & = & \frac{1}{2}[J_{0}(z_{1})J_{0}(z_{2})\omega_{0}-2\tilde{A}_{0}\cos(\Delta t+\delta\phi_{21})]\sigma_{z}\nonumber \\
 &  & +\sum_{j=1}^{2}\frac{\tilde{A}_{j}}{4}[{\rm e}^{{\rm i}(\omega_{j}t+\phi_{j})}\sigma_{-}+{\rm e}^{-{\rm i}(\omega_{j}t+\phi_{j})}\sigma_{+}],
\end{eqnarray}
where
\begin{equation}
\tilde{A}_{0}=\omega_{0}J_{1}(z_{1})J_{1}(z_{2}),
\end{equation}
\begin{equation}
\Delta=\omega_{2}-\omega_{1},
\end{equation}
\begin{equation}
\delta\phi_{21}=\phi_{2}-\phi_{1},
\end{equation}
\begin{equation}
\tilde{A}_{j}=2A_{j}(1-\xi_{j}).
\end{equation}

The second part $H_{2}^{\prime}(t)=H^\prime(t)-H^\prime_{\rm CHRW}(t)$ contains
the faster oscillatory terms including the higher beat-frequency terms and is given by
\begin{widetext}
\begin{eqnarray}
H_{2}^{\prime}(t) & = & \omega_{0}\sigma_{y}\sum_{k=1}^{\infty}\sum_{n=1}^{2k}J_{n}(z_{1})J_{2k+1-n}(z_{2})\left\{ \sin[n\theta_{1}+(2k+1-n)\theta_{2}]\right.\nonumber\\
 &  & \left.-(-1)^{n}\sin[n\theta_{1}-(2k+1-n)\theta_{2}]\right\} \nonumber \\
 &  & +\omega_{0}\sigma_{y}\sum_{n=2}^{\infty}\left\{ J_{2n-1}(z_{1})J_{0}(z_{2})\sin[(2n-1)\theta_{1}]\right.\nonumber\\
 &  & \left.+J_{0}(z_{1})J_{2n-1}(z_{2})\sin[(2n-1)\theta_{2}]\right\} \nonumber \\
 &  & +\omega_{0}\sigma_{z}\sum_{n=1}^{\infty}[J_{2n}(z_{1})J_{0}(z_{2})\cos(2n\theta_{1})+J_{0}(z_{1})J_{2n}(z_{2})\cos(2n\theta_{2})]\nonumber \\
 &  & +\omega_{0}\sigma_{z}\sum_{k=2}^{\infty}\sum_{n=1}^{2k-1}(-1)^{n}J_{n}(z_{1})J_{2k-n}(z_{2})\cos[n\theta_{1}-(2k-n)\theta_{2}]\nonumber \\
 &  & +\omega_{0}\sigma_{z}\sum_{k=1}^{\infty}\sum_{n=1}^{2k-1}J_{n}(z_{1})J_{2k-n}(z_{2})\cos[n\theta_{1}+(2k-n)\theta_{2}].
\end{eqnarray}
\end{widetext}
In spite of the complexity, this part may be reasonably neglected under
certain conditions. Specifically, when $A_j/\omega_j\sim1$, one has $z_j\sim1$ because of $\xi_j\sim1$. On recalling the properties of the Bessel functions of the first kind, we note that the oscillating amplitudes in $H_{2}^\prime(t)$ become considerably small. On the other hand, roughly speaking, the fast-oscillating terms of $H_{2}^{\prime}(t)$ are responsible for the higher-order multiphoton processes that $n+m$ ($m=3,5,7...$) photons of one component of the bichromatic field are absorbed and $n$ photons of the other component are emitted. Such processes become important for very large driving amplitudes and the near- or on-resonance $\omega_0\approx m\omega_j$. Therefore, we neglect the contribution of $H_{2}^\prime(t)$ and retain $H_{{\rm CHRW}}^{\prime}(t)$ as the effective Hamiltonian. $H_{{\rm CHRW}}^{\prime}(t)$ is referred to as the CHRW Hamiltonian and is expected to take account of the effects of the counter-rotating couplings via the renormalized parameters although it takes a similar form as the RWA Hamiltonian.

To proceed, we transform $H_{{\rm CHRW}}^{\prime}(t)$ into a frame rotating at
the frequency $\omega_{1}$ by the rotation transformation with $R(t)=\exp[{\rm i}(\omega_{1}t+\phi_{1})\sigma_{z}/2]$,
yielding
\begin{eqnarray}
\tilde{H}_{{\rm CHRW}}(t) & = & \frac{1}{2}\left[\tilde{\delta}_{1}-2\tilde{A}_{0}\cos(\Delta t+\delta\phi_{21})\right]\sigma_{z}+\frac{\tilde{A}_{1}}{4}\sigma_{x}\nonumber \\
 &  & +\frac{\tilde{A}_{2}}{4}[{\rm e}^{{\rm i}(\Delta t+\delta\phi_{21})}\sigma_{-}+{\rm e}^{-{\rm i}(\Delta t+\delta\phi_{21})}\sigma_{+}],\label{eq:Heff}
\end{eqnarray}
where
\begin{equation}
\tilde{\delta}_{1}=\omega_{0}J_{0}(z_{1})J_{0}(z_{2})-\omega_{1}.
\end{equation}
Importantly, the effective Hamiltonian is now periodic in time with
a period $T=2\pi/|\Delta|$. In other words, the CHRW method transforms the aperiodic Hamiltonian with an external bichromatic  field to a periodic one.

\subsection{Floquet theory and transition probability}

We calculate the time-evolution operator for $\tilde{H}_{{\rm CHRW}}(t)$ by using the Floquet theory, which states that the evolution
operator takes the formal form~\cite{Shirley_1965}:
\begin{equation}
\tilde{U}(t,t_{0})=\sum_{\gamma=\pm}{\rm e}^{-{\rm i}\tilde{\varepsilon}_{\gamma}(t-t_{0})}|\tilde{u}_{\gamma}(t)\rangle\langle\tilde{u}_{\gamma}(t_{0})|,
\end{equation}
where $|\tilde{u}_{\gamma}(t)\rangle$ is the Floquet state and has
the same periodicity as $\tilde{H}_{{\rm CHRW}}(t)$ and $\tilde{\varepsilon}_{\gamma}$
is the real-valued quasienergy. The index $\gamma$ denotes two linearly independent Floquet states. They satisfy the following equation:
\begin{equation}
[\tilde{H}_{{\rm CHRW}}(t)-{\rm i}\partial_{t}]|\tilde{u}_{\gamma}(t)\rangle=\tilde{\varepsilon}_{\gamma}|\tilde{u}_{\gamma}(t)\rangle.\label{eq:eigenEq}
\end{equation}
Solving the above differential equations can be transformed into an
eigenvalue problem in linear algebra. To this end, we introduce the
so-called Sambe space spanned by the Floquet bases $\{|\uparrow\rangle\otimes|n\rangle,|\downarrow\rangle\otimes|n\rangle|n=0,\pm1,\pm2,\ldots\}$,
where $|\uparrow\rangle$ and $|\downarrow\rangle$ are the eigenstates
of $\sigma_{z}$ with the eigenvalues $+1$ and $-1$, respectively,
and $|n\rangle\equiv\exp({\rm i}n\Delta t)$ is a Fourier basis and is defined as an infinite sparse vector with one nonvanishing component that equals 1 in the $n$th position numbered from a specified origin~\cite{Sambe_1973}.
The inner product for the Fourier bases is defined as $\langle n|m\rangle=\frac{1}{T}\int_{0}^{T}{\rm e}^{-{\rm i}n\Delta t+{\rm i}m\Delta t}dt=\delta_{n,m}.$
In terms of the extended Hilbert space, equation~(\ref{eq:eigenEq}) is converted into a time-independent
matrix equation
\begin{equation}
\tilde{{\cal H}}_{{\rm CHRW}}|\tilde{u}_{\gamma}\rangle=\tilde{\varepsilon}_{\gamma}|\tilde{u}_{\gamma}\rangle,
\end{equation}
where $\tilde{{\cal H}}_{{\rm CHRW}}=\tilde{H}_{{\rm CHRW}}(t)-{\rm i}\partial_{t}$
is the Floquet Hamiltonian and $|\tilde{u}_{\gamma}\rangle$ is a vector with the components
given by the Fourier coefficients of $|\tilde{u}_{\gamma}(t)\rangle$. The explicit form of the Floquet Hamiltonian reads
\begin{eqnarray}\label{eq:Hfchrw}
\tilde{{\cal H}}_{{\rm CHRW}} & = & \frac{1}{2}\left(\tilde{\delta}_{1}\sigma_{z}+\frac{\tilde{A}_{1}}{2}\sigma_{x}\right)\otimes I+\sigma_0\otimes\sum_{n=-\infty}^{\infty}n\Delta|n\rangle\langle n|\nonumber \\
 &  & +\frac{1}{4}\sum_{n=-\infty}^{\infty}(\tilde{A}_{2}\sigma_{-}-2\tilde{A}_{0}\sigma_{z}){\rm e}^{{\rm i}\delta\phi_{21}}|n+1\rangle\langle n|\nonumber \\
 &  & +\frac{1}{4}\sum_{n=-\infty}^{\infty}(\tilde{A}_{2}\sigma_{+}-2\tilde{A}_{0}\sigma_{z}){\rm e}^{-{\rm i}\delta\phi_{21}}|n\rangle\langle n+1|,
\end{eqnarray}
where $\sigma_0$  and $I$ are the $2 \times 2$ and infinite-size identity matrices, respectively.
Although $\tilde{{\cal H}}_{{\rm CHRW}}$ is an infinite-size matrix,
it can be numerically diagonalized with an appropriate
truncation. On diagonalizing $\tilde{{\cal H}}_{{\rm CHRW}}$, one
obtains the quasienergies and the Fourier coefficients of the Floquet
states. It is sufficient to choose the quasienergies in the first
Brillouin zone $(-|\Delta|/2,|\Delta|/2]$ and the corresponding Floquet
states, which completely determine the time-evolution operator for
the effective Hamiltonian $\tilde{H}_{{\rm CHRW}}(t)$.

With $\tilde{U}(t,t_0)$ at hand, we can obtain the time evolution operator for the
original Hamiltonian, which is related to the former via
\begin{equation}
U(t,t_{0})={\rm e}^{-S(t)}R^{\dagger}(t)\tilde{U}(t,t_{0})R(t_{0}){\rm e}^{S(t_{0})}.
\end{equation}
Based on $U(t,t_{0})$, we can calculate the transient transition
probability of finding the qubit in the excited state at
time $t$ when it is in the ground state at time $t_{0}$,
\begin{eqnarray}
P(t,t_{0}) & = & |\langle\uparrow|U(t,t_{0})|\downarrow\rangle|^{2}\nonumber \\
 & = & {\rm Tr}[|\uparrow\rangle\langle\uparrow|U(t,t_{0})|\downarrow\rangle\langle\downarrow|U^{\dagger}(t,t_{0})]\nonumber \\
 & = & {\rm Tr}[R(t){\rm e}^{S(t)}|\uparrow\rangle\langle\uparrow|{\rm e}^{-S(t)}R^{\dagger}(t)\tilde{U}(t,t_{0})\nonumber \\
 &  & \times {\rm e}^{S(t_{0})}R(t_{0})|\downarrow\rangle\langle\downarrow|{\rm e}^{-S(t_{0})}R^{\dagger}(t_{0})\tilde{U}^{\dagger}(t,t_{0})]\nonumber \\
 & = & \frac{1}{2}-\frac{1}{4}\sum_{\mu,\nu=z,\pm}f_{\mu}(t)f_{\nu}(t_{0})\sum_{\lambda,\gamma=\pm}{\rm e}^{-{\rm i}(\tilde{\varepsilon}_{\gamma}-\tilde{\varepsilon}_{\lambda})(t-t_{0})}\nonumber \\
 &  & \times\langle\tilde{u}_{\lambda}(t)|\sigma_{\mu}|\tilde{u}_{\gamma}(t)\rangle\langle\tilde{u}_{\gamma}(t_{0})|\sigma_{\nu}|\tilde{u}_{\lambda}(t_{0})\rangle,
\end{eqnarray}
where
\begin{equation}
f_{z}(t)=\cos\left[\sum_{j=1}^{2}z_{j}\sin(\omega_{j}t+\phi_{j})\right],
\end{equation}
\begin{equation}
f_{\pm}(t)=\mp {\rm i}\sin\left[\sum_{j=1}^{2}z_{j}\sin(\omega_{j}t+\phi_{j})\right]{\rm e}^{\pm {\rm i}(\omega_{1}t+\phi_{1})}.
\end{equation}

The time-averaged transition probability can be given by
\begin{equation}\label{eq:pav}
  \overline{P}=\overline{P(t,t_{0})}=\frac{1}{2}(1-d^2),
\end{equation}
\begin{eqnarray}\label{eq:d}
  d& = & \sum_{n=-\infty}^{\infty}\left[J_{n}(z_{1})J_{-n}(z_{2}){\rm e}^{-{\rm i}n\delta\phi_{21}}X_{++,n}^{z}\right.\nonumber\\
  & &+J_{n+1}(z_{1})J_{-n}(z_{2}){\rm e}^{{\rm i}n\delta\phi_{21}}X_{++,-n}^{+}\nonumber \\
 &  & \left.+J_{n+1}(z_{1})J_{-n}(z_{2}){\rm e}^{-{\rm i}n\delta\phi_{21}}X_{++,n}^{-}\right],
\end{eqnarray}
where the overline indicates the average over time and
\begin{equation}
X_{++,n}^{\mu}=\frac{1}{T}\int_{0}^{T}\langle\tilde{u}_{+}(t)|\sigma_{\mu}|\tilde{u}_{+}(t)\rangle {\rm e}^{-{\rm i}n\Delta t}dt.
\end{equation}
From Equation~(\ref{eq:pav}), one readily notes that the behavior of $\overline{P}$ totally depends on $d$. It is important to understand the properties of $d$.
First, it is simple to prove that $d$ is real-valued by using the relations $X^{z}_{++,-n}=[X^{z}_{++,n}]^\ast$ and $X^{+}_{++,-n}=[X^{-}_{++,n}]^\ast$. Second, although the phase difference $\delta\phi_{21}$ appears in~(\ref{eq:d}), it is found via numerical calculation that $d$ is independent of the phases and so does $\overline{P}$. This is consistent with the previous finding that the phases of the bichromatic field are found to have no influence on the quasienergies~\cite{Potvliege_1992,Poertner_2020}. For this reason, we consider $\phi_1=\phi_2=0$ throughout this work. Third,
it is evident that when $d=0$, $\overline{P}$ takes on the maximal value 1/2, which signifies the resonance response
of the qubit to the external field. Therefore, the resonance conditions are related to the zeros of $d$. In addition, we note that the role of $d$ is actually equivalent to the derivative of the quasienergy of the GFT with respect to $\omega_0$ by comparing Equation~(\ref{eq:pav}) with~(\ref{eq:gftpav}).

\begin{figure}
\centering
  \includegraphics[width=\columnwidth]{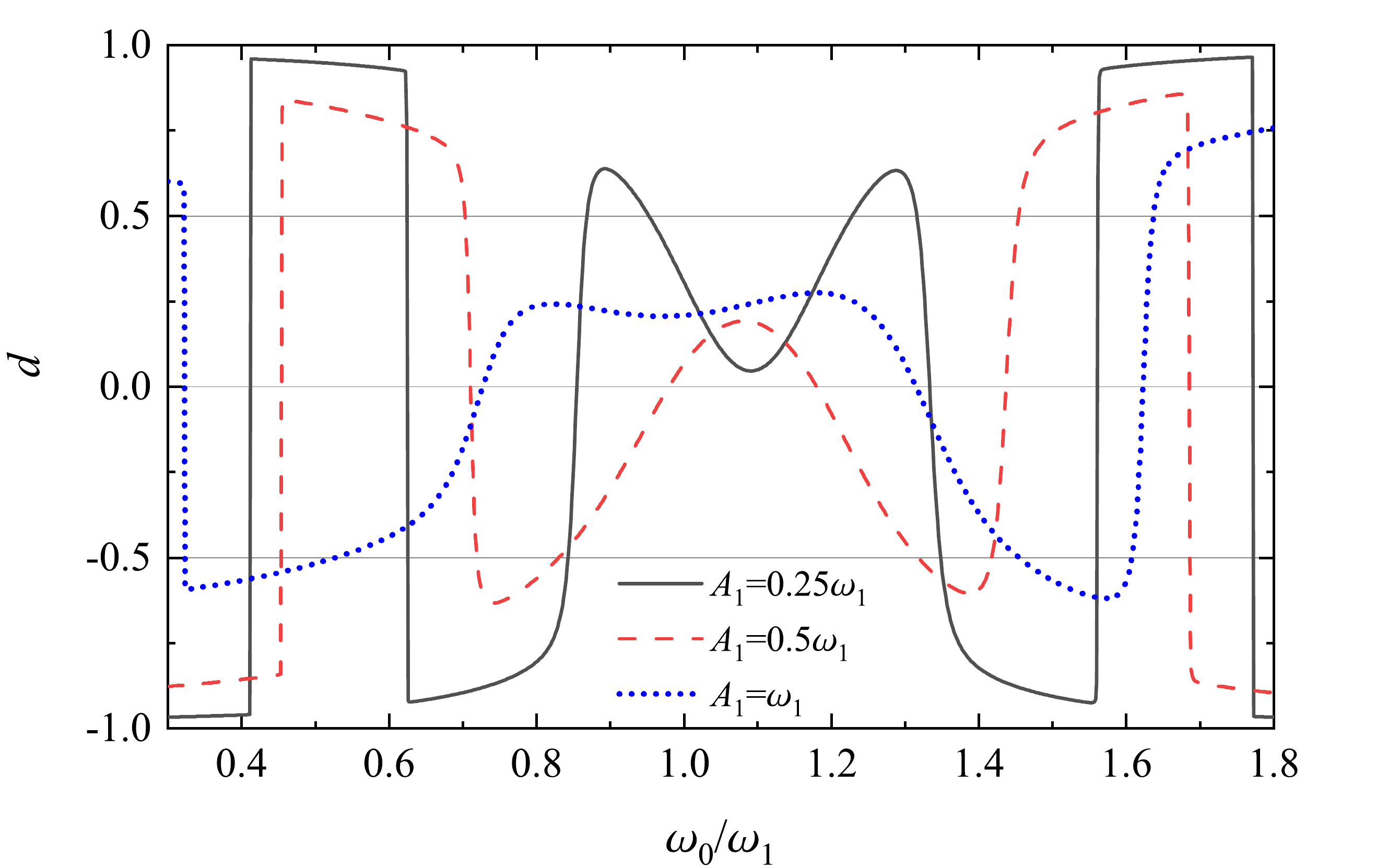}
  \caption{Dimensionless quantity $d$ in Equation~(\ref{eq:d}) versus $\omega_0$ for $r=1$, $\Delta=0.2\omega_1$, and the three values of $A_1$.}\label{fig1}
\end{figure}

From the above analysis, we note that the resonance frequency can be simply obtained by solving $d=0$ for variable $\omega_0$ with the other parameters being fixed. \textbf{Figure~\ref{fig1}} shows the typical behavior of $d$ as a function of $\omega_0$ for $r=1$, $\Delta=0.2\omega_1$, and the three values of $A_1$. We set $\omega_1$ as units hereafter. Interestingly, we find that in general the sign of $d$ changes around its zeros as the variation of $\omega_0$. This property allows us to efficiently find the zeros of $d$, namely, the resonance positions, by using the bisection algorithm.

\begin{figure*}
  \includegraphics[width=2\columnwidth]{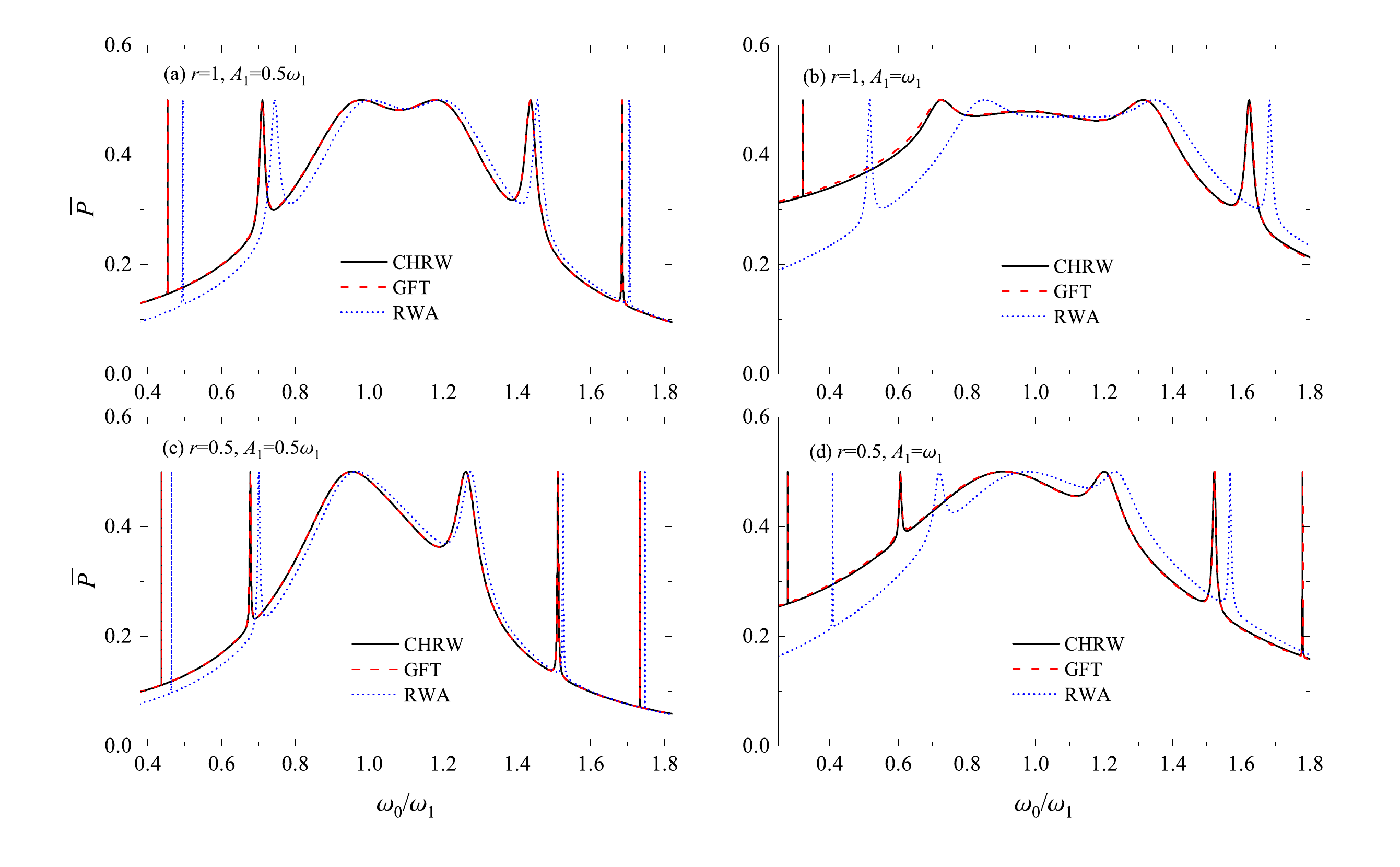}
  \caption{Time-averaged transition probability $\overline{P}$ versus $\omega_0$ for $\Delta=0.2\omega_1$, two values of $r$, and two values of $A_1$.}\label{fig2}
\end{figure*}

For comparison and to distinguish the effects of the counter-rotating couplings from the CHRW Hamiltonian, let us revisit the widely used RWA Hamiltonian. For the present bichromatically driven qubit, the RWA Hamiltonian is obtained by neglecting the counter-rotating couplings $\sum_{j=1}^{2}\frac{A_{j}}{4}[{\rm e}^{{\rm i}(\omega_{j}t+\phi_{j})}\sigma_{+}+{\rm e}^{-{\rm i}(\omega_{j}t+\phi_{j})}\sigma_{-}]$ in~(\ref{eq:Ham}). Similarly, one can transform the RWA Hamiltonian into the frame rotating at frequency $\omega_1$, yielding
\begin{eqnarray}
\tilde{H}_{{\rm RWA}}(t)& = & \frac{1}{2}(\omega_{0}-\omega_{1})\sigma_{z}+\frac{A_{1}}{4}\sigma_{x}+\frac{A_{2}}{4}[{\rm e}^{{\rm i}(\Delta t+\delta\phi_{21})}\sigma_{-}\nonumber\\
&  &+{\rm e}^{-{\rm i}(\Delta t+\delta\phi_{21})}\sigma_{+}].\label{eq:HRWA}
\end{eqnarray}
Clearly, the transformed RWA Hamiltonian is periodic in time and thus the Floquet
theory can be applied. In the previous works, the RWA Hamiltonian is usually treated in the frame rotating at the average frequency $\overline{\omega}=(\omega_1+\omega_2)/2$~\cite{Ruyten_1989,Ruyten_1992,Agarwal_1991,Ficek_1996}. Interestingly, in such a frame, the mathematical form of the RWA Hamiltonian becomes the same as that of the Rabi model when $A_{1}=A_{2}$, i.e., $r=1$. Nevertheless, for the numerical calculation, Equation~(\ref{eq:HRWA}) is preferred since the convergence is much faster. Although the RWA leads to the simplified treatment based on the Floquet theory, it neglects the counter-rotating couplings which induce resonance shifts and complicated beat behavior.

\section{Results and discussions}\label{sec:results}

\begin{figure*}
  \includegraphics[width=\columnwidth]{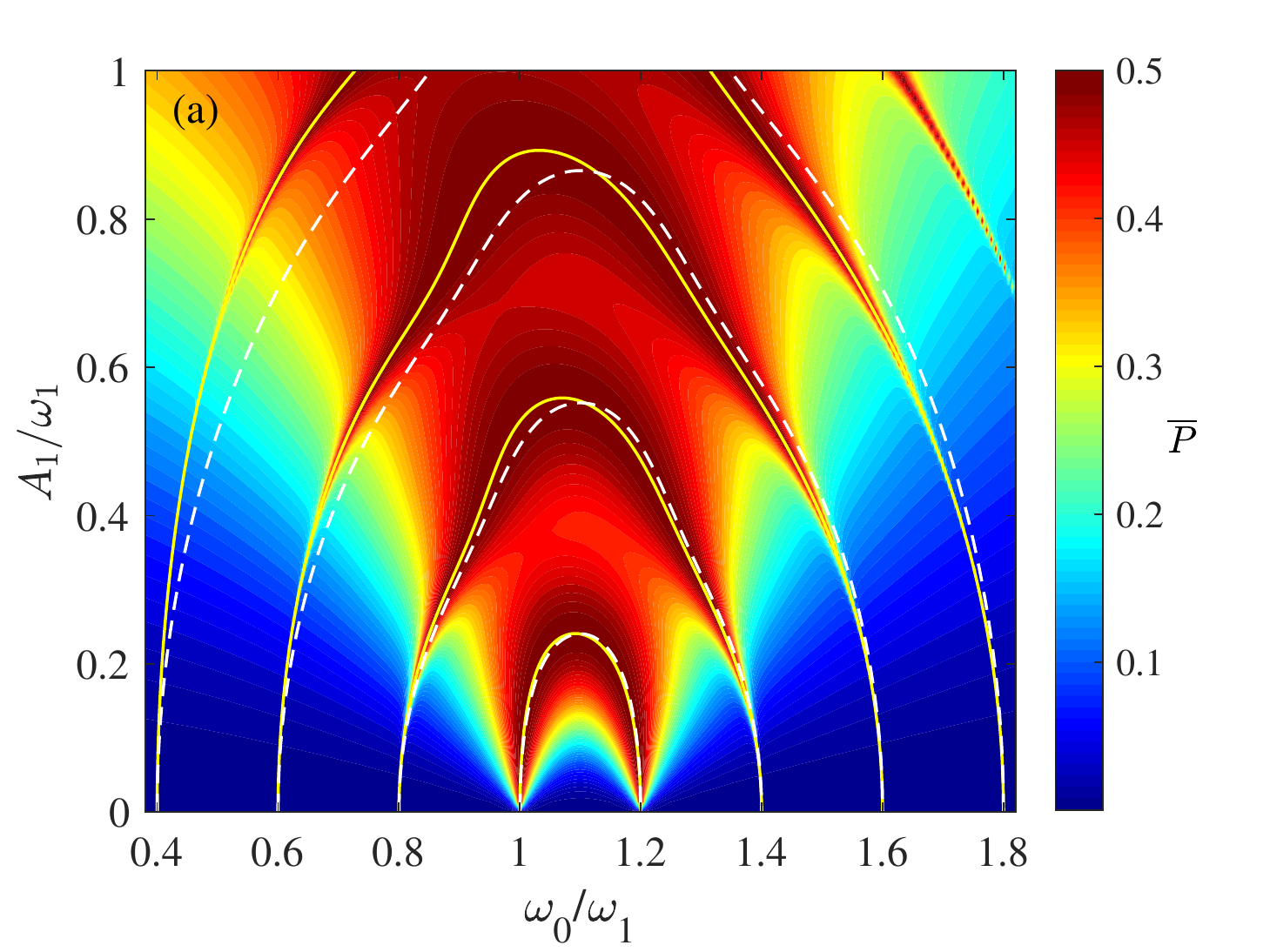}
  \includegraphics[width=\columnwidth]{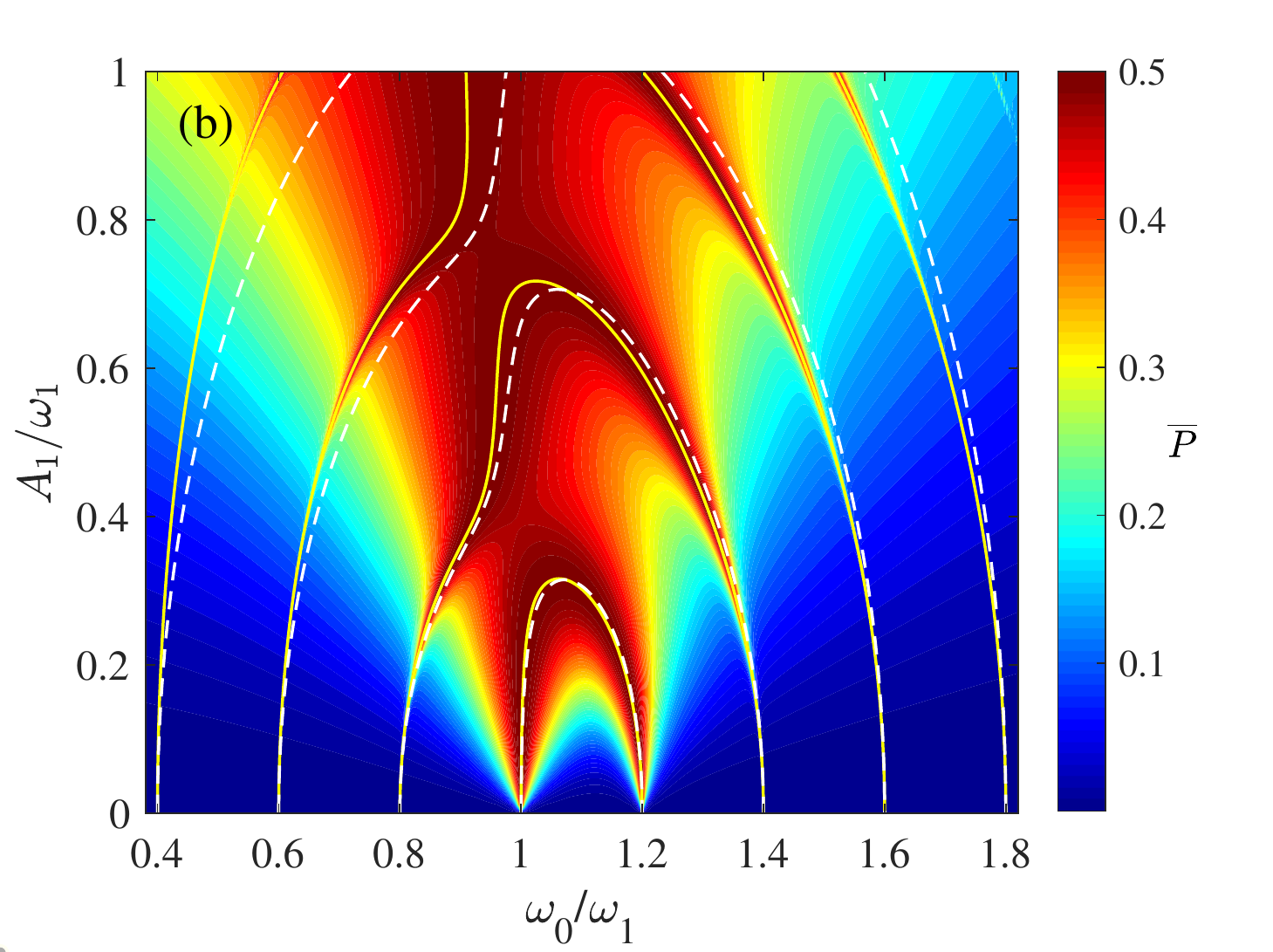}
  \caption{Contour plot of time-averaged transition probability $\overline{P}$ versus $\omega_0$ and $A_1$ calculated from the GFT method for $\Delta=0.2\omega_1$ and two values of $r$. (a) $r=1$, (b) $r=0.5$. The solid lines are the CHRW results and indicate the resonance positions ($\overline{P}=1/2$). The dashed lines are the RWA resonance positions.}\label{fig3}
\end{figure*}

In this section, we study the resonance behaviors  of the bichromatically driven qubit by using the CHRW, RWA, GFT, and RK methods. The codes for generating the data can be found in ~\cite{Yangit}.

\subsection{High beat-frequency case}

We first consider a relatively high beat-frequency case, i.e., $|\Delta|\sim0.1\omega_1$. In this case, the GFT method is chosen to be the benchmark, which is numerically exact. The details of the GFT method is presented in Appendix. In the following, we first examine the performance of the CHRW and RWA methods and then study the Bloch-Siegert shifts as well as their influence on multiphoton dynamical processes.

To examine the accuracy of the CHRW and the RWA methods on the predictions of the resonance, we calculate the time-averaged transition probability $\overline{P}$ as a function of $\omega_0$ for the fixed $\omega_1$ and $\omega_2$. The difference of the two frequencies is set as $\Delta=0.2\omega_1$. \textbf{Figure~\ref{fig2}} shows the resonance curves obtained from the three methods for the two values of $r$ and the two values of $A_1$.  We find that the CHRW results (solid lines) are in good agreement with the GFT results (dashed lines) for both values of $r$ even if $A_1/\omega_1=1$. In fact,
it is straightforward to numerically verify that for the considered values of $A_1$ and $r$, the CHRW method is also accurate when the value of $|\Delta|$ is neither too large nor too small, i.e., $|\Delta|/\omega_1\sim0.1$. The present findings lead to the conclusion that the CHRW Hamiltonian is capable of predicting accurate resonance positions of the bichromatically driven qubit when $A_1/\omega_1\sim1$, $r\sim1$, $|\Delta|\sim0.1\omega_1$, and $\omega_0\sim\omega_1$. On the other hand, when comparing the RWA results with the CHRW results, one readily finds that they are inconsistent on three aspects. First, there are shifts between the RWA and CHRW resonance peaks. Obviously, these shifts result from the counter-rotating couplings. In this sense, they are similar to the Bloch-Siegert shift found in the Rabi model~\cite{Bloch_1940,Shirley_1965}. Therefore, the shifts between the RWA and the CHRW peaks are referred to as the Bloch-Siegert shifts. In addition, Figure~\ref{fig2} indicates that the larger the driving amplitudes are, the larger the shifts between the RWA and CHRW resonance peaks become. Second, the RWA resonance curve is symmetric about the average frequency $\overline{\omega}$ ($\overline{\omega}=1.1\omega_1$ in the present case) provided $r=1$ while the CHRW and GFT curves do not. The symmetry of the RWA result can be simply attributed to the fact that the RWA Hamiltonian has the same mathematical form as the Rabi model in the frame rotating at the average frequency $\overline{\omega}$ as long as $r=1$. Clearly, if $r\neq1$, namely, the intensities of the two components of the bichromatic field are not equal, one finds that the RWA curves are also apparently asymmetric as those of the CHRW and GFT. Third, the RWA resonance width may be quite different from those of CHRW and GFT [see the widths of resonance peaks near $\omega_0=0.4\omega_1$ in Figure~\ref{fig2}(b)]. This reflects that the counter-rotating coupling has significant influence on the multiphoton resonance width.

Next, we move to illustrate the Bloch-Siegert shifts as the variation of the driving amplitudes, which reflects the deviation between the RWA and exact resonance positions. To this end, we first calculate the time-averaged transition probability as a function of $\omega_0$ and $A_{1}$ by using the GFT method for $\Delta=0.2\omega_1$ and two values of $r$. \textbf{Figure \ref{fig3}} shows the contour plots of the GFT numerical results, providing the insights into the resonance positions and widths of the bichromatically driven qubit. It is clear to see that in the $\omega_0$-$A_1$ plane, the resonance peaks form the separated resonance bands, i.e., the resonance exhibits band structure. These bands correspond to either single-photon or multiphoton processes, which can be characterized by the resonance positions in the weak-driving limit, namely, the endpoints of the band. In general, the band that connects $\omega_0=(n+1)\omega_1-n\omega_2$ and $\omega_0=(n+1)\omega_2-n\omega_1$ as $A_1\rightarrow0$ is the $(2n+1)$-photon resonance band, where the qubit absorbs $n+1$ photons from one component of the bichromatic field and emits $n$ photons of the other component. Although $2n+1$ photons participate at these resonances, there is only one net photon absorbed. Besides, the maxima $(\overline{P}=1/2)$ at the bands are the resonance positions. In the band gaps, the resonance is forbidden.

We use the CHRW and RWA methods to calculate the resonance positions of the four bands ranging from the single- to seven-photon resonance, which are represented by solid lines and dashed lines in Figure~\ref{fig3}, respectively. When $A_1/\omega_1>0.2$, there are significant Bloch-Siegert shifts between the RWA and CHRW resonance positions at the multiphoton resonance bands.
However, when $A_{1}/\omega_{1}<0.2$, the RWA resonance positions agree with the CHRW ones, indicating that the Bloch-Siegert shifts are negligible and the RWA may be a good approximation when the driving is sufficiently weak. The present results suggest that the Bloch-Siegert shifts become notable at the multiphoton resonance bands and in the strong-driving regime.

\begin{figure}
\centering
  \includegraphics[width=\columnwidth]{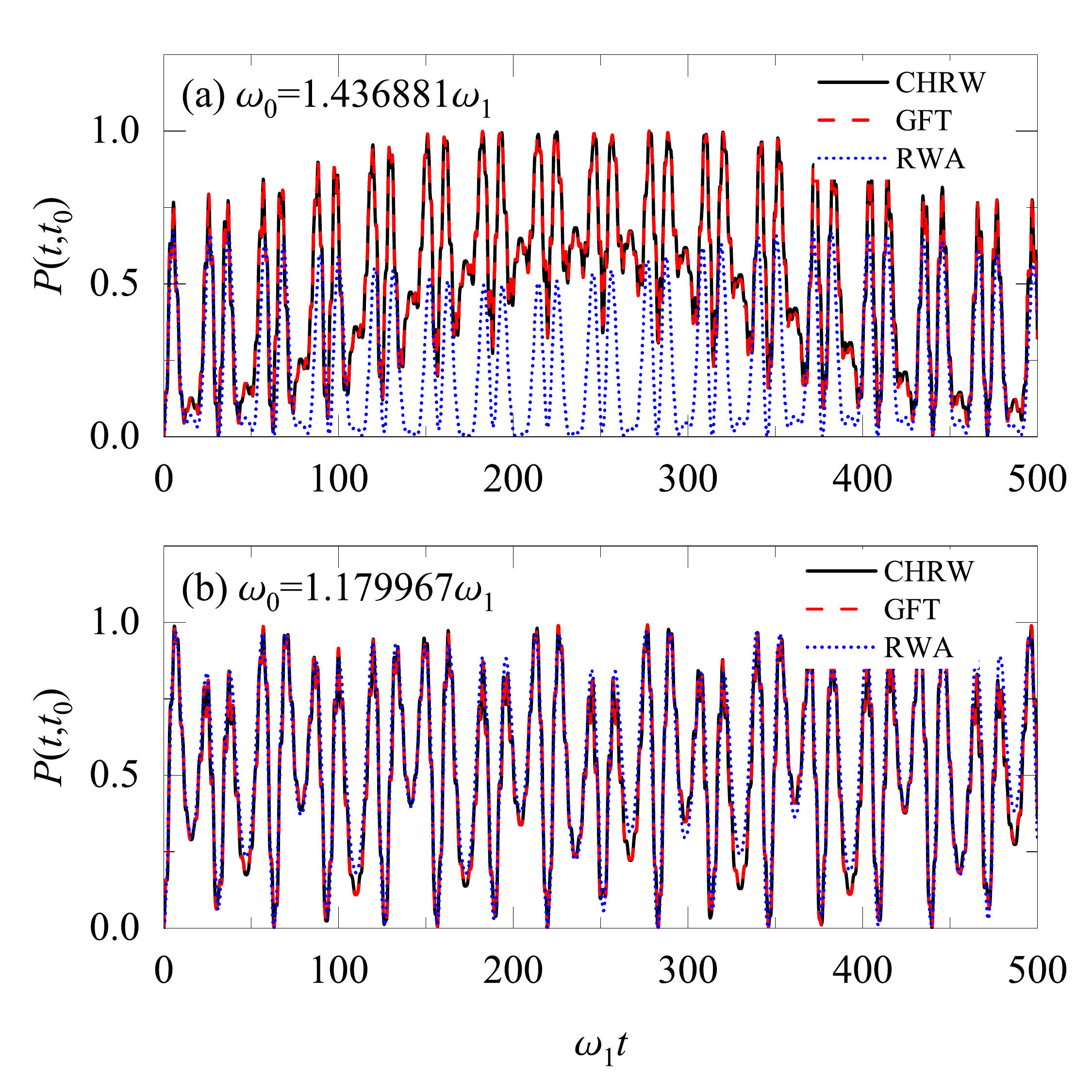}
  \caption{Transient transition probability $P(t,t_0)$ versus $\omega_1 t$ for $t_0=0$, $A_1=0.5\omega_1$, $r=1$, $\Delta=0.2\omega_1$, and two values of $\omega_0$. (a) $\omega_0=1.436881\omega_1$ is a five-photon resonance frequency. (b) $\omega_0=1.179967\omega_1$ is a three-photon resonance frequency.}\label{fig4}
\end{figure}

We now illustrate the influence of the Bloch-Siegert shifts on the dynamics of the bichromatically driven qubit. As is known, at the resonance where the magnitude of Bloch-Siegert shift is comparable with the resonance width, the transient transition probability can be dramatically different whether the RWA is used or not~\cite{Yan_2017}. In Figure~\ref{fig3}, we see that such situation may occur at the multiphoton resonance band. In \textbf{Figure~\ref{fig4}} we use the three methods to calculate the transient transition probability $P(t,t_0)$ as a function of $t$ for $t_0=0$, $A_1=0.5\omega_1$, $r=1$, and two values of $\omega_0$. We first consider $\omega_0=1.436881\omega_1$, which corresponds to the abscissa of the maximum of the CHRW curve near $\omega_0=1.4\omega_{1}$ in Figure~\ref{fig2}(a) and is a five-photon resonance frequency [see Figure~\ref{fig3}(a)]. Figure~\ref{fig4}(a) shows that the RWA dynamics is completely different from the CHRW dynamics while the latter agrees with the GFT. Such difference can be attributed to the fact that the Bloch-Siegert shift leads to that the resonance frequency of the CHRW Hamiltonian becomes a ``far'' off-resonant frequency of the RWA Hamiltonian when the Bloch-Siegert shift quantifying the detuning between the RWA and CHRW resonance frequencies is comparable with the RWA resonance width~\cite{Yan_2017}. Moreover, one can verify that the essential difference between the RWA and CHRW (GFT) dynamics arises at other CHRW resonance positions as long as the Bloch-Siegert shift is comparable with the resonance width. On the contrary, if the Bloch-Siegert shift is much smaller than the resonance width, one can expect that there is no essential difference between the RWA and CHRW (GFT) dynamics. To verify this, we consider $\omega_0=1.179967\omega_1$, which is the abscissa of the maximum of the CHRW peak near $\omega_0=1.2\omega_{1}$ in Figure~\ref{fig2}(a) and is a three-photon resonance frequency. Figure \ref{fig4}(b) shows that there is no essential difference between the RWA and CHRW (GFT) dynamics. Nevertheless, we note that the RWA method is unable to capture accurate beat behavior as the CHRW method. The present results confirm that the Bloch-Siegert shift can cause essential difference between the RWA and CHRW (GFT) in the present bichromatically driven qubit under the multiphoton resonance condition where the Bloch-Siegert shift is comparable to the resonance width. In addition, the CHRW method is found to predict not only  accurate resonance positions but also accurate dynamics.

\begin{figure}
\centering
  \includegraphics[width=\columnwidth]{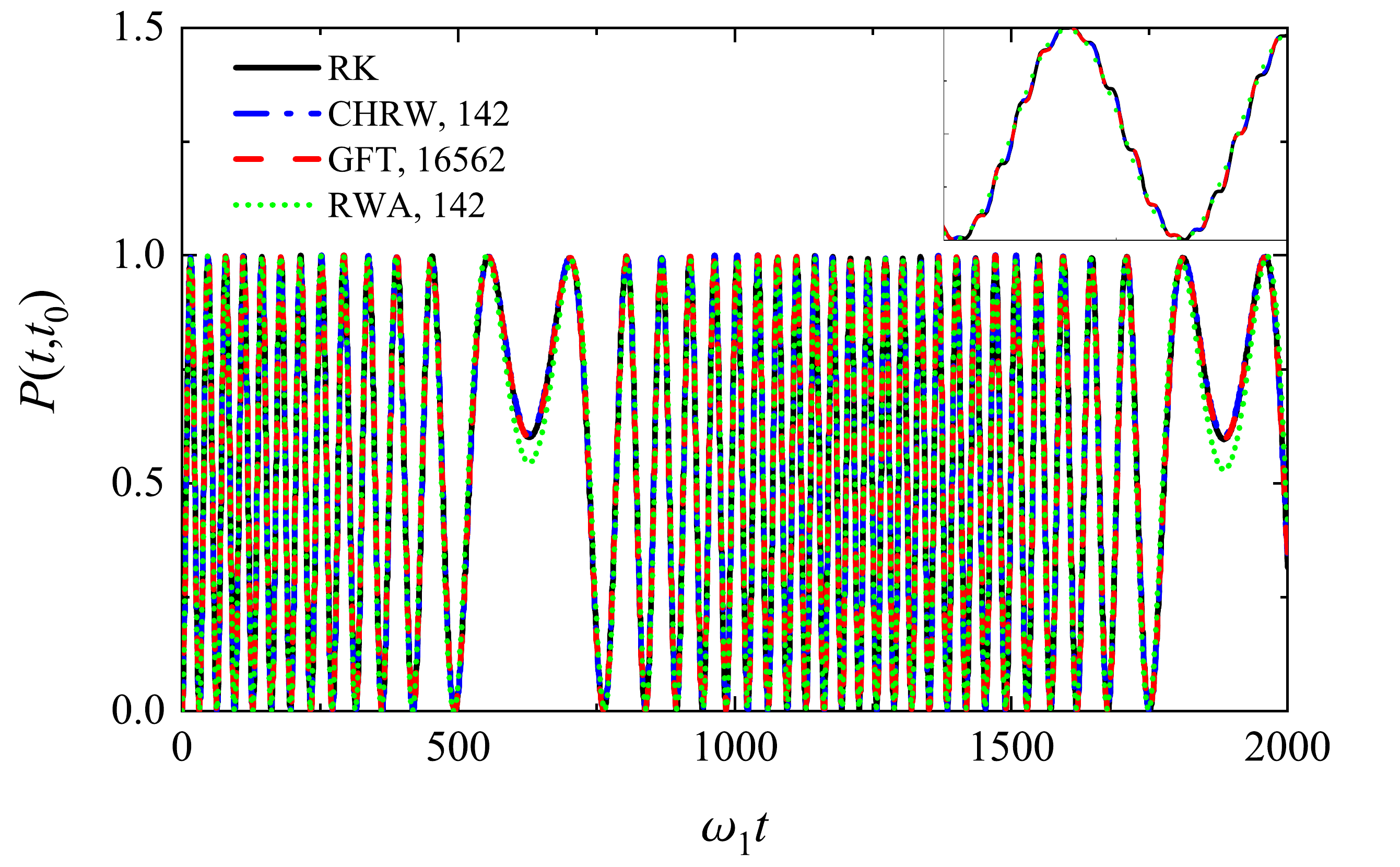}
  \caption{Transient transition probability $P(t,t_0)$ versus $\omega_1 t$ calculated from the RK, CHRW, GFT, and RWA methods for $t_0=0$, $A_1=0.2\omega_1$, $r=1$, $\Delta=0.005\omega_1$, and $\omega_0=\omega_1$. The inset shows the zoom of curves in the interval $[1350,1400]$. The number in the legends indicates the dimension of the truncated Floquet or generalized Floquet matrices.}\label{fig5}
\end{figure}

\subsection{Low beat-frequency case}
In this section, we discuss the performance of the CHRW method in the relatively low beat-frequency cases, i.e., $|\Delta|\ll\omega_1$. In such cases, we run into difficulty to examine the accuracy of the CHRW method by comparing its time-averaged transition probability with that of the GFT. The GFT approach is found to be difficult to converge when $|\Delta|\ll\omega_1$, $r\sim1$, and $\omega_0\sim\omega_1$. Consequently, we compare the transient transition probability calculated by the CHRW method with that of the RK method, which is used to directly integrate the time-dependent Schr\"{o}dinger equation.
We find that the CHRW method is valid even in the vanishingly small beat-frequency case. We summarize the valid regimes of the CHRW method for the four magnitudes of $|\Delta|$ in Table~\ref{tab1}. It turns out that the smaller $|\Delta|$ is, the smaller amplitudes the CHRW is valid for. Besides, the size of the truncated Floquet matrix of the CHRW Hamiltonian also increases for convergence as $|\Delta|$ decreases. Typically, when $|\Delta|\sim0.1\omega_1$, a $62\times62$ truncated Floquet matrix is sufficient to guarantee the convergence. However, when $|\Delta|\sim10^{-4}\omega_1$, a $2082\times2082$ truncated Floquet matrix is needed for convergence. Nevertheless, we find that the CHRW method has great advantage over the numerically exact GFT approach in such low beat-frequency regimes.

To exemplify the advantage of the CHRW method, in \textbf{Figure~\ref{fig5}}, we show the transient transition probabilities calculated from the RK, CHRW, GFT, and RWA methods for $A_1=0.2\omega_1$, $r=1$, $\Delta=0.005\omega_1$, and $\omega_0=\omega_1$. We emphasis that even for $A_1=0.2\omega_1$, a $16562\times16562$ truncated generalized Floquet matrix is needed to get the accurate dynamics and it takes a few hours CPU time in a modern PC. On the contrary, in the CHRW treatment, a $142\times142$ truncated Floquet matrix is sufficient to get the converged and accurate result and it just takes few seconds CPU time. The CHRW is not only as efficient as the RWA treatment but also captures the correct beat behaviors missed in the latter. In general, one finds that the smaller $|\Delta|$ is, the more difficult the GFT approach is to converge even for a relatively small amplitude and $r\sim1$, which can be attributed to the poor convergence of the two-mode Fourier series used in the GFT approach.

\begin{figure}
\centering
  \includegraphics[width=\columnwidth]{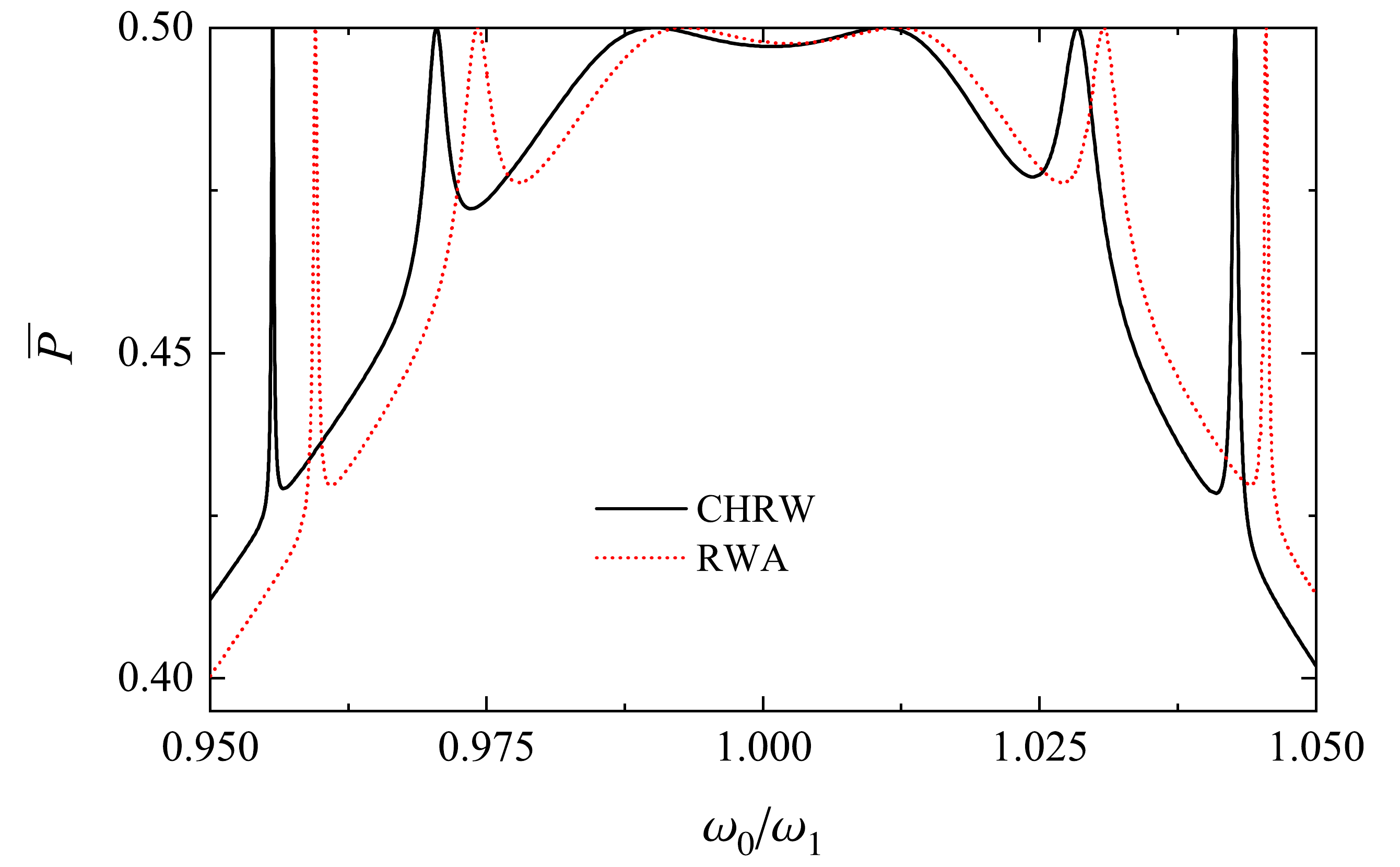}
  \caption{Time-averaged transition probability $\overline{P}$ versus $\omega_0$ for $\Delta=0.005\omega_1$,  $A_1=0.2\omega_1$, and $r=1$.}\label{fig6}
\end{figure}

To show the Bloch-Siegert shift in the low beat-frequency case, we calculate the time-averaged transition probabilities by using the CHRW and RWA methods.
\textbf{Figure~\ref{fig6}} shows $\overline{P}$ as a function of $\omega_0$ for $\Delta=0.005\omega_1$, $A_1=0.2\omega_1$, and $r=1$. We note that the line shapes of $\overline{P}$ in Figure~\ref{fig6} are similar as those in Figure~\ref{fig2}(a).
Besides, the spacing between the resonance peaks in the small $\Delta$ case is less than that in the large $\Delta$ case. In other words, for the small $\Delta$, the resonance bands become dense in the $\omega_0$-$A_1$ plane. This is not favorable for studying the resonance shifts.

\begin{table}
\caption{Valid regimes of the CHRW method for $\omega_{0}\sim\omega_{1}$,
$r\sim1$, and various values of $\Delta$.}\label{tab1}
\begin{ruledtabular}
\begin{tabular}{cccc}
$|\Delta|\sim10^{-1}\omega_{1}$ & $|\Delta|\sim10^{-2}\omega_{1}$ & $|\Delta|\sim10^{-3}\omega_{1}$ & $|\Delta|\sim10^{-4}\omega_{1}$\tabularnewline
$A_{1}\sim\omega_{1}$ & $A_{1}\sim0.3\omega_{1}$ & $A_{1}\sim0.2\omega_{1}$ & $A_{1}\sim0.1\omega_{1}$\tabularnewline
\end{tabular}
\end{ruledtabular}
\end{table}

\section{Conclusions}
In summary, we have studied the resonance and dynamics of a qubit strongly driven by a bichromatic field beyond the RWA. A periodic CHRW Hamiltonian has been derived from the aperiodic original Hamiltonian based on the unitary transformations and it enables us to accurately calculate the main resonance positions and dynamics by making use of the Floquet theory. When the beat frequency is relatively large, in comparison with the GFT results, we have shown that the CHRW method provides an accurate description of the main resonance and dynamics of the bichromatically driven qubit over a strong-driving regime where the RWA breaks down, suggesting that the effects of the counter-rotating couplings of the bichromatic field have been properly incorporated in the CHRW Hamiltonian in its valid regime. Besides, the CHRW method has a much higher efficiency in numerical calculation than the GFT method. Using the CHRW, RWA, and GFT methods, we have illustrated the Bloch-Siegert shifts induced by the counter-rotating couplings of the bichromatic field. Such shifts become notable at the multiphoton resonance bands and in the strong driving regimes. Moreover, we found that a situation where the magnitude of the Bloch-Siegert shift becomes comparable with the resonance width can occur at the multiphoton resonance band. In such a situation, the RWA and CHRW theories yield essentially different dynamics at the CHRW resonance positions because of the Bloch-Siegert shift.
When the beat frequency is relatively small, we find that the CHRW is capable of efficiently predicting accurate results for the amplitudes comparable with the driving frequency while the GFT approach becomes difficult to converge.

The present CHRW method offers insights
into the effects of the counter-rotating couplings
of the bichromatic field on the resonance and dynamics
in the strong-driving regime. In addition, it may be useful in the quantum battery researches with the strong bichromatic field, similar to the monochromatic case~\cite{Zhang_2019,Chen_2020,Crescente_2020}.

\appendix*
\section{Generalized Floquet theory}
According to the generalized Floquet theory, the time-evolution operator for the bichromatically driven two-level system under study takes the form~\cite{Chu_2004}
\begin{equation}
U(t,t_{0})=\sum_{\gamma=\pm}|u_{\gamma}(t)\rangle\langle u_{\gamma}(t_{0})|{\rm e}^{-{\rm i}\varepsilon_{\gamma}(t-t_{0})},
\end{equation}
where $\varepsilon_\gamma$ is a real-valued quasienergy and $|u_{\gamma}(t)\rangle$ possesses a two-mode Fourier expansion,
\begin{equation}\label{eq:uttmf}
|u_{\gamma}(t)\rangle=\sum_{n,m=-\infty}^{\infty}{\rm e}^{{\rm i}(n\omega_{1}+m\omega_{2})t}|u^{(n,m)}_{\gamma}\rangle.
\end{equation}
From the time-dependent Schr\"{o}dinger equation, one simply obtains that $\varepsilon_\gamma$ and $|u_\gamma(t)\rangle$ satisfy the following equation:
\begin{equation}\label{eq:eiggft1}
[H(t)-{\rm i}\partial_{t}]|u_{\gamma}(t)\rangle=\varepsilon_{\gamma}|u_{\gamma}(t)\rangle.
\end{equation}

In the present formalism, the main task is to calculate the quasienergies and the two-mode Fourier coefficients of the unknown vectors $|u_\gamma(t)\rangle$. To this end, we expand
the Hamiltonian in terms of two-mode Fourier series,
\begin{eqnarray}\label{eq:htmf}
H(t) & = & \sum_{n,m=-\infty}^{\infty}H^{(n,m)}{\rm e}^{{\rm i}(n\omega_{1}+m\omega_{2})t}\nonumber \\
 & = & H^{(0,0)}+H^{(1,0)}{\rm e}^{{\rm i}\omega_{1}t}+H^{(-1,0)}{\rm e}^{-{\rm i}\omega_{1}t}\nonumber \\
 &  & +H^{(0,1)}{\rm e}^{{\rm i}\omega_{2}t}+H^{(0,-1)}{\rm e}^{-{\rm i}\omega_{2}t},
\end{eqnarray}
where
\begin{equation}
  H^{(0,0)}=\frac{1}{2}\omega_{0}\sigma_{z},
\end{equation}
\begin{equation}
  H^{(\pm1,0)}=\frac{1}{4}A_{1}{\rm e}^{\pm {\rm i}\phi_{1}}\sigma_{x},
\end{equation}
\begin{equation}
H^{(0,\pm1)}=\frac{1}{4}A_{2}{\rm e}^{\pm {\rm i}\phi_{2}}\sigma_{x},
\end{equation}
and $H^{(n,m)}=0$ otherwise. To proceed, we substitute Equations (\ref{eq:uttmf}) and (\ref{eq:htmf}) into (\ref{eq:eiggft1}), we derive the equations for the two-mode Fourier
coefficients and quasienergy,
\begin{equation}
\sum_{n,m=-\infty}^{\infty}[H^{(k-n,l-m)}+(n\omega_{1}+m\omega_{2})\delta_{n,k}\delta_{l,m}]|u_{\gamma}^{(n,m)}\rangle=\varepsilon_{\gamma}|u_{\gamma}^{(k,l)}\rangle.
\end{equation}
These equations can be reformulated in a matrix form
\begin{equation}
  {\cal H}_{F2}|u_{\gamma}\rangle=\varepsilon_{\gamma}|u_{\gamma}\rangle.
\end{equation}
Here, ${\cal H}_{F2}$ is the two-mode Floquet Hamiltonian and is
given by
\begin{eqnarray}
{\cal H}_{F2} & = & \frac{1}{2}\omega_{0}\sigma_{z}\otimes I\otimes I+\sigma_{0}\otimes\sum_{n=-\infty}^{\infty}n\omega_{1}|n\rangle\langle n|\otimes I\nonumber \\
 &  & +\sigma_{0}\otimes I\otimes\sum_{m=-\infty}^{\infty}m\omega_{2}|m\rangle\langle m|\nonumber \\
 &  & +\frac{A_{1}}{4}{\rm e}^{-{\rm i}\phi_{1}}\sigma_{x}\otimes\sum_{n=-\infty}^{\infty}|n\rangle\langle n+1|\otimes I\nonumber \\
 &  & +\frac{A_{1}}{4}{\rm e}^{{\rm i}\phi_{1}}\sigma_{x}\otimes\sum_{n=-\infty}^{\infty}|n+1\rangle\langle n|\otimes I\nonumber \\
 &  & +\frac{A_{2}}{4}{\rm e}^{-{\rm i}\phi_{2}}\sigma_{x}\otimes I\otimes\sum_{m=-\infty}^{\infty}|m\rangle\langle m+1|\nonumber \\
 &  & +\frac{ A_{2}}{4}{\rm e}^{{\rm i}\phi_{2}}\sigma_{x}\otimes I\otimes\sum_{m=-\infty}^{\infty}|m+1\rangle\langle m|,
\end{eqnarray}
$|u_{\gamma}\rangle$
is a column vector whose components are the two-mode Fourier coefficients
$|u_{\gamma}^{(n,m)}\rangle$. With appropriate truncation, we can
numerically diagonalize the two-mode Floquet Hamiltonian to obtain
the quasienergies and the corresponding eigenvectors $|u_{\gamma}\rangle$.
These can be used to calculate the transient and time-averaged transition probabilities~\cite{Ho_1984,Chu_2004}
\begin{equation}
 P(t,t_{0})=\left|\sum_{k,l=-\infty}^{\infty}{\rm e}^{{\rm i}(k\omega_{1}+l\omega_{2})t}\langle\uparrow,k,l|{\rm e}^{-{\rm i}{\cal H}_{F2}(t-t_{0})}|\downarrow,0,0\rangle\right|^{2},
\end{equation}
\begin{eqnarray}
\overline{P}&=&\sum_{k,l,n,m=-\infty}^{\infty}\sum_{\gamma=\pm}|\langle\uparrow,k,l|u_{\gamma,n,m}\rangle\langle u_{\gamma,n,m}|\downarrow,0,0\rangle|^{2}\nonumber\\
                  &=&\frac{1}{2}\left[1-4\left(\frac{\partial \varepsilon_\gamma}{\partial \omega_0}\right)^{2}\right],\label{eq:gftpav}
\end{eqnarray}
where $|u_{\gamma,n,m}\rangle$ is the eigenvector of ${\cal H}_{F2}$
associated with the shifted quasienergy $\varepsilon_{\gamma,n,m}\equiv\varepsilon_{\gamma}+n\omega_{1}+m\omega_{2}$
and $|\uparrow(\downarrow),k,l\rangle\equiv|\uparrow(\downarrow)\rangle\otimes|k\rangle\otimes|l\rangle$.


\begin{acknowledgments}
Support from the National Natural Science Foundation of China (Grants No. 12005188 and No. 11774226)
is gratefully acknowledged.
\end{acknowledgments}

\bibliography{bichromaticdrive}


\end{document}